\documentclass[fleqn,usenatbib]{mnras}

\usepackage{newtxmath,newtxtext}
\usepackage{siunitx}
\usepackage{pythonhighlight}
\usepackage[noabbrev]{cleveref}

\usepackage[T1]{fontenc}

\DeclareRobustCommand{\VAN}[3]{#2}
\let\VANthebibliography\thebibliography
\def\thebibliography{\DeclareRobustCommand{\VAN}[3]{##3}\VANthebibliography}

\usepackage{graphicx}	
\usepackage{bm}

\newcommand{\err}[3]{$#1^{+ #2}_{- #3}$}

\title[Dark energy model comparison]{Non-Gaussian Likelihoods for Type Ia Supernovae Cosmology: Implications for Dark Energy and $H_0$}

\author[T. Lovick et al.]{
Toby Lovick,$^{1}$
Suhail Dhawan,$^{1}$
Will Handley $^{1,2}$
\\
$^{1}$ Kavli Institute for Cosmology, University of Cambridge, Madingley Road, Cambridge CB3 0HA, UK \\
$^{2}$ Astrophysics Group, Cavendish Laboratory University of Cambridge,
JJ Thomson Avenue, Cambridge, CB3 0HE, UK
}

\date{4 December 2023}

\pubyear{0000}

\begin{document}
\label{firstpage}
\pagerange{\pageref{firstpage}--\pageref{lastpage}}
\maketitle
\begin{abstract}
The latest improvements in the scale and calibration of Type Ia supernovae catalogues allow us to constrain the specific nature and evolution of dark energy through its effect on the expansion history of the universe. We present the results of Bayesian cosmological model comparison on the SNe~Ia catalogue Pantheon+, where Flat $\Lambda$CDM is preferred by the data over all other models and we find moderate evidence ($\Delta \log \mathcal{Z} \sim 2.5$) to reject a number of the alternate dark energy models. The effect of peculiar velocity corrections on model comparison is analysed, where we show that removing  the peculiar velocity corrections results in a varying fit on non-$\Lambda$CDM parameters. As well as comparing cosmological models, the Bayesian methodology is extended to comparing the scatter model of the data, testing for non-gaussianity in the Pantheon+ Hubble residuals. We find that adding a scale parameter to the Pantheon+ covariances, or alternately using a multivariate Student's t-distribution fits the data better than the fiducial analysis, producing a cosmology independent evidence increase of $\Delta \log \mathcal{Z} = 2.29 $ and $2.46$ respectively. This improved treatment of the scatter decreases the uncertainty in the  constraint on the Hubble constant, finding $H_0 = 73.67 \pm 0.99 $ \si{km.s^{-1}.Mpc^{-1}}, in  $ 5.7 \sigma$ tension with Planck. We also explore $M_B$ transition models as a potential solution for the Hubble tension, finding no evidence to support these models among the SNe data.

\end{abstract}




\section{Introduction}
Type Ia supernovae (SNe~Ia) are extremely bright thermonuclear explosions of a carbon-oxygen white dwarf in a binary system \citep[see][for a review of their observational properties]{Maguire2017}. After correcting for the lightcurve shape \citep{phillips1993}, colour \citep{tripp1998} and host galaxy properties \citep[e.g.][]{Kelly2010,Sullivan_2010}, they have a small dispersion in their brightness, making them excellent distance indicators for cosmology \citep[e.g., see][for a review of their cosmological utility]{Goobar2011}. They were instrumental in the discovery of universal accelerated expansion \citep{Riess1998,Perlmutter_1999} and are crucial for precision measurements of dark energy properties, e.g. energy density, equation of state and changes with cosmic time \citep{brout_2022}. 


Observations of the SN~Ia magnitude-redshift relation, as well as high-precision early and late universe probes, e.g. the cosmic microwave background power spectrum \citep{Planck2018}, baryon acoustic oscillations \citep{Bourboux2020}, and the distribution and evolution of galaxies \citep{d_Amico_2020}, have led to the inception of a standard cosmological model, termed $\Lambda$CDM. It is a model with only 6 parameters, explaining a wide range of datasets and phenomena. However, there exist significant theoretical and observational problems with dark energy that are unsolved by $\Lambda$CDM \citep{Perivolaropoulos_2022}, which motivates looking into physics beyond our standard cosmology. The most relevant problem with $\Lambda$CDM for this paper is the Hubble tension, the observed $5 \sigma$ tension between the CMB-measured value of $H_0 \sim 67.4 \pm 0.5$ \si{km.s^{-1}.Mpc^{-1}} \citep{Planck2018}, and the distance measure Hubble constant, constrained on the Pantheon+ data as $H_0 = 73.04 \pm 1.04$ \si{km.s^{-1}.Mpc^{-1}} \citep{Riess2022}. Tensions (Hubble or otherwise) in our current theory motivate searching for physics beyond the standard cosmology; alternate formulations of dark energy.

Despite the mathematical simplicity of a cosmological constant there is no reason a priori to assume that dark energy takes this form, and many alternative theories exist that also function as an accelerating factor in the dynamics of the scale of the universe. The simplest modification is $w$CDM, where the equation of state parameter $P = w\rho$ is no longer fixed at $w = -1$ for dark energy. Other models allow for a dynamical dark energy, where instead of acting like a cosmological constant, the accelerating power of dark energy evolves with cosmological time. A final category of models allows modifications to General Relativity; since our evidence for dark energy comes from Einstein's equations, it's possible to justify the appearance of an accelerating universe via a novel theory of gravity. The list of models considered in this paper are drawn largely from \cite{Dhawan2020} and cover a wide range of physical motivations and theoretical approaches to modelling dark energy. 

Pantheon+ is the culmination of a series of efforts to standardise the luminosity of SNe~Ia, to improve their constraining power as a cosmological probe. The data have been corrected in a number of ways, accounting for statistical and systematic effects on the data: see \cite{brout_2022} Figure 1 for a full list of the companion papers that form the total catalogue. All of these effects lead to the corrected apparent magnitudes in the Pantheon+ catalogue, with their intrinsic scatter reduced to $\sim 0.1$ mag.

The usage of Bayesian techniques in cosmology has greatly increased in recent years \citep{hobson2010bayesian}. The Bayesian approach is to update our current belief, the prior distribution, using the observed data and its likelihood to form a new belief about a hypothesis, the posterior distribution. This allows us to take an uninformative prior (such as $\Omega_m \sim U[0,1]$) and update it using the supernova data to infer the value of parameters within a dark energy model. Beyond parameter inference, we can update our level of belief in competing dark energy models, as in \cite{Kurek_2008}, \cite{Handley_2021}. Starting from the belief that all dark energy models are equally likely, we can calculate the posterior probabilities of competing models as they are favoured by the data.

A review of the relative merits and demerits of a Bayesian model selection analysis are explored in detail in \cite{bernado2009,lindley2000,bayesinthesky}. For comparing various different models, a key advantage is that the Bayes factor, i.e. the logarithm of the ratio of the Bayesian evidence, penalises models with larger degrees of freedom if the additional degrees do not lead to a significant improvement in the quality of the fit  \citep{Hergt_2021}. This effectively applies an ``Occam's razor" in the analysis. In this paper, we move beyond just dark energy model comparison and use the Bayes factor to not only compare different cosmological models using SNe~Ia data but also different likelihood models and how well they describe the Pantheon+ data. We present our methodology in \cref{sec:method}, our results in \cref{sec:results} and discuss them in light of the current literature in \cref{sec:discussion}. Our conclusions are presented in \cref{sec:conclusion}

\section{Methodology}\label{sec:method}
\subsection{Bayes' Theorem} \label{Bayes}
All equations in this section are contained in \cite{bayesinthesky}. \\

In the context of parameter inference and model comparison, Bayes' theorem says that the distribution of a set of parameters $\theta$ within a model $\mathcal{M}$ gets updated by data $d$ as:

\begin{equation}
    p(\theta|d,\mathcal{M}) = \frac{p(d|\theta,\mathcal{M}) p(\theta|\mathcal{M})}{p(d|\mathcal{M})},
    \label{posterior}
\end{equation}
or 
\begin{equation*}
    \text{Posterior} = \frac{\text{Likelihood}\times\text{Prior}}{\text{Evidence}} \equiv \mathcal{P} = \frac{\mathcal{L}\pi}{\mathcal{Z}}.
\end{equation*}
In this analysis we take the prior distribution of models to be uniform, i.e. 
\begin{equation}
    p(\mathcal{M}) = \frac{1}{\text{Number of models}},
\end{equation}
so that model comparison is quantified by the evidence ratio: 
\begin{equation}
    \frac{p(\mathcal{M}_0|d)}{p(\mathcal{M}_1|d)} = \frac{p(d|\mathcal{M}_0)}{p(d|\mathcal{M}_1)} = B_{01},
    \label{Bayes Factor Equation}
\end{equation} 
where $B_{01}$ is the Bayes factor. We compute these evidences by integrating out the posterior distribution over parameter space in \cref{posterior} to find
\begin{equation}
    p(d|\mathcal{M}) = \int_{\Omega_\theta} p(d|\theta,\mathcal{M}) p(\theta|\mathcal{M}) d\theta = \langle \mathcal{L} \rangle _{\pi}.
    \label{EvidenceEquation}
\end{equation}
From here on the evidence of a model is labelled with $\mathcal{Z}$, and all model comparisons are expressed in terms of $\Delta \log \mathcal{Z} = \log(\mathcal{Z}_0/\mathcal{Z}_1)$. \cref{table:jeff} shows the Jeffreys' Scale \citep{jeffreys1961} which we use to judge the significance of our results.
\begin{table}
\centering
\caption{The odds, probabilities, and interpretations of various Bayes factor as given by \protect\cite{jeffreys1961}. Here the probabilities are the posterior probability of the superior model, starting with an even prior $p(\mathcal{M}_0) = p(\mathcal{M}_1) = 0.5$.}

\hspace{5ex} \\

\begin{tabular}{@{\extracolsep{\fill}} l c c c}

\hline 
$|\log B_{01}|$ & Odds & Probability & Strength of evidence \\ [0.5ex] 

\hline 
$<1.0$ & $\lesssim 3:1$ & $<0.750$ & Inconclusive \\
$1.0$ & $\sim 3:1$ & $0.750$ & Weak  \\
$2.5$ & $\sim 12:1$ & $0.923$& Moderate  \\
$5$ & $\sim 150:1$ & $0.993$ & Strong  \\ [1ex]  
\hline 

\end{tabular}

\label{table:jeff}
\end{table}
\subsection{Occam Penalty} \label{occam}
As mentioned previously Bayesian model comparison has a built in Occam's razor term, that works to rule out over-complicated models. This takes the form of Kullback–Leibler divergence \citep{kullback}, the statistical distance between the posterior and prior :
\begin{equation}
    \mathcal{D}_\textrm{KL}(\theta|\mathcal{M}) = \int p(\theta | d,\mathcal{M}) \log \frac{p(\theta | d,\mathcal{M})}{p(\theta | \mathcal{M})}d\theta = \int \mathcal{P} \log \frac{\mathcal{L}}{\mathcal{Z}}d\theta.
\end{equation}
Splitting up the logarithm demonstrates how this term penalises the evidence \citep{Hergt_2021}:
\begin{equation}
    \log \mathcal{Z} = \langle \log\mathcal{L} \rangle _{\mathcal{P}} - \mathcal{D}_\textrm{KL},
    \label{DKLequation}
\end{equation}
so the evidence of a model is equal to the goodness of fit, as measured by the posterior averaged log likelihood, minus a penalty term, the KL-divergence. $\Delta \log \mathcal{Z}$ can be dissected even further, using the approximation:
\begin{equation} \label{laplaceapprox}
	\langle \log \mathcal{L}\rangle_\mathcal{P} \approx \log \mathcal{L}_{\text{max}} -\frac{n}{2},
\end{equation}
where $n$ is the model dimension. This expression is exact when the likelihood is a multivariate Gaussian in $\theta$. This can be generalised by replacing $n$ with $\hat{d}$,the Bayesian Model Dimensionality \citep{complexity}, which is a measure of the effective number of constrained parameters in a theory. This holds for parameters with approximately Gaussian or uniform posteriors, i.e. over ``well-behaved" parameter spaces. Substituting this into \cref{DKLequation}:
\begin{equation} \label{occampenaltyapprox}
	\Delta \log \mathcal{Z} \approx \Delta(\log \mathcal{L}_{\text{max}} ) -  \frac{\Delta\hat{d}}{2}  - \Delta(\mathcal{D}_\textrm{KL}).
\end{equation}
The evidence difference is (approximately) the best possible fit a model can find, plus a $\mathcal{D}_\textrm{KL}$ penalty, plus a dimensionality penalty. 

\subsection{Cosmological Model Fitting}

Given some parameters $\theta$ in a given cosmology, we calculate luminosity distances with
\begin{equation}
\label{DL}
    d_L(z) = \frac{c(1+z)}{H_0\sqrt{|\Omega_k|}} \text{sinn} \left(\sqrt{|\Omega_k|} \int ^{z}_0 \frac{dz'}{E(z';\theta)}\right),
\end{equation}
where $E^2(z) = H^2(z)/H^2_0$ is the dimensionless Hubble parameter, which relates the expansion history to cosmological parameters. This is related to the distance modulus $\mu$ by:
\begin{equation}
    \mu(z) = 5\log(\frac{d_L(z)}{10\text{pc}}).
    \label{moduli}
\end{equation}
We now calculate Hubble residuals, the deviation of the observed distances to the model:
\begin{equation}
    \Delta_i = \mu_i - \mu_{\text{model}}(z_i) = (m_{B\text{ } i}-M_B) -\mu_{\text{model}}(z_i).
\end{equation}
The corrected $m_B$ values are given in Pantheon+ as:
\begin{equation}
    m_B^{corr} = m_B + \alpha x_1 - \beta c - \delta_{\text{bias}} +\delta_{\text{host}}.
    \label{mbcorr}
\end{equation}
This equation standardises across light curves via the Tripp parameters $\alpha,\beta$ \citep{tripp1998}, as well as correcting for host galaxy/survey specific selection biases \citep{brout_2022}.

\begin{figure}
    \centering
  \includegraphics{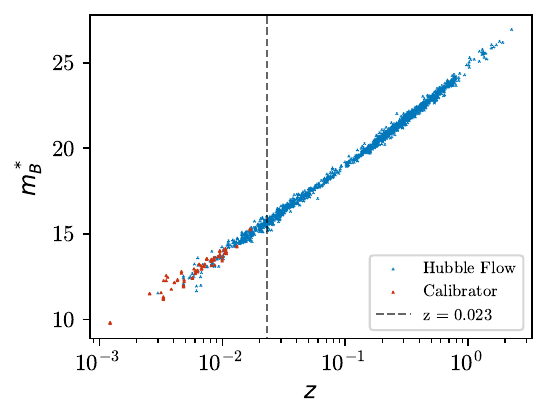}
  \caption{The redshift-apparent magnitude data in the Pantheon+ catalogue, where here $m_B^*$ refers to the corrected Pantheon+ magnitudes. The dashed grey line shows the redshift cutoff applied to the data, at $z = 0.023$. This data is drawn from the ``zHD" and ``m\textunderscore b\textunderscore corr" columns of Pantheon+.}
  \label{hubblediagram}
\end{figure}

For Calibrator observations (SNe that have an independent distance measure from Cepheid Variable stars), residuals are calculated as:
\begin{equation} \label{cephres}
    \Delta = (m_B-M_B)-\mu_{\text{ceph}}.
\end{equation}  
This cosmology-independent term only constrains $M_B$, which propagates through the data so $H_0$ can be estimated on the remaining ``Hubble Flow" SNe. This data is shown in \cref{hubblediagram}, along with the redshift cut-off applied to the Hubble Flow SNe data at $z = 0.023$ (as implemented in the SH$_0$ES calculation of $H_0$ \citep{Riess2022}). We then construct a Gaussian likelihood:
\begin{equation}
    \log\mathcal{L} = -\frac{1}{2}\left( \Delta^T \bm{C}^{-1} \Delta + \log |2\pi \bm{C}| \right),
    \label{gaussianlikelihood}
\end{equation}
where $\bm{C}$ is the covariance matrix of the Pantheon+ data. The process of defining this matrix is detailed in Section 2.2 of \cite{Dhawan2020}. This likelihood function, as well as the prior distributions detailed in \cref{table:prior} are fed into the nested sampling software \texttt{PolyChord} \citep{polychord,Polychord2}. While nested sampling is a numerical method, all errors in the evidence values are $< 0.05$, so are quoted/plotted without error bars.

\subsection{Dark energy Models}\label{DEmodels}
This section describes the set of dark energy models considered in this paper, with brief explanations of the motivation and mechanism of each model. The priors used for each model parameter are given in \cref{table:prior}. From here on the fiducial 6-parameter cosmology is referred to as Flat/F$\Lambda$CDM, to distinguish it from $\Lambda$CDM with non-zero curvature.

\subsection*{$\Lambda$CDM and $w$CDM}
All of the models considered in this paper act as modifications to F$\Lambda$CDM, and due to the many successful predictions of F$\Lambda$CDM must behave significantly like it to be worth considering. Despite F$\Lambda$CDM being a 6-parameter model, many of the parameters do not directly affect the expansion history, so as a dark energy model it has a single parameter $\Omega_m = 1-\Omega_\Lambda$. 

The first 4 models of this paper: F$\Lambda$CDM, $\Lambda$CDM, $w$CDM, and Flat $w$CDM (F$w$CDM) treat dark energy as having a static equation of state throughout cosmic evolution. $\Lambda$CDM allows for an over or under-dense universe $\Omega_m + \Omega_\Lambda \neq 1$, which we parameterise as curvature density $\Omega_m + \Omega_\Lambda = 1 - \Omega_k$. As well as affecting the evolution of the Hubble parameter, the presence of curvature also affects luminosity distance geometrically, as on a large scale the universe is no longer Euclidean \citep{cosmologicalmodels}. 

$w$CDM is a phenomenological model where the equation of state is not fixed to -1 (the value corresponding to a cosmological constant) but is instead a free parameter. We test this model both as a flat cosmology, and with the curvature density as a free parameter.

These four models are nested in one another; starting from curved $w$CDM setting $w = -1, \Omega_k = 0$, or both leads to $\Lambda$CDM, F$w$CDM and F$\Lambda$CDM respectively. The dimensionless Hubble Parameter for $w$CDM is 
\begin{equation}
    \frac{H^2}{H^2_0} = \Omega_M(1+z)^3 +\Omega_k(1+z)^2 + (1-\Omega_M - \Omega_k)(1+z)^{3(1+w)}.
\end{equation}
\subsection*{One-Parameter Slow Roll Dark Energy}
Scalar field dark energy theories model current day cosmic acceleration and early time cosmic inflation with the same mechanism; a scalar field rolling down a potential. \cite{SLepian2014} showed that, for a potential satisfying the ``Slow-Roll Conditions" that allow it to behave like dark energy, the generic behaviour of this form of dark energy is independent of the initial value and shape of the potential. This generality makes it a highly desirable model, as it has only a single additional parameter $\delta w_0$ in its Hubble parameter expression:
\begin{equation}
    \frac{H^2}{H^2_0} = \Omega_M(1+z)^3 + (1-\Omega_M)\left( \frac{(1+z)^3}{\Omega_M(1+z)^3+1-\Omega_M}\right)^{\frac{\delta w}{1-\Omega_M}}.
\end{equation}
\subsection*{Growing Neutrino Mass}
This model aims to address the ``cosmological coincidence" problem; a model that proposes an interaction with the scalar field that halts its evolution can justify why $\Omega_m \sim \Omega_\Lambda$, if the interaction only happens at a typical matter/energy density. In the case of this growing neutrino mass model \citep{Fardon2004,Wetterich2007}, the scalar field stops evolving due to neutrinos becoming non-relativistic (at a value of $\rho_\Lambda$ typical of an epoch transition such as is observed today). This model introduces two additional parameters, the early dark energy density $\Omega_e$, and the current day neutrino density $\Omega_\nu$.

In this cosmology the dark energy density has different late/early time behavior:
\begin{equation*}
\Omega_\textrm{de}
\begin{cases}
\frac{\Omega_{de}a^3+2\Omega_\nu (a^{3/2}-a^3)}{1-\Omega_{de}(1-a^3)+2\Omega_\nu (a^{3/2}-a^3)} & \text{; } a > a_t \\ \\
\Omega_e & \text{; } a < a_t.
\end{cases}
\end{equation*}
The Hubble expression is then: 
\begin{equation}
    \frac{H^2}{H^2_0} = \frac{\Omega_m (1+z)^3}{1-\Omega_{de}(a)}.
\end{equation}
\subsection*{Algebraic Thawing}
Thawing models contain a dark energy with $w = -1$ at high redshift, that departs from this value over cosmic time to less negative values. One such way of realising this evolution is Algebraic Thawing \citep{Linder_2007}, a dark energy that evolves in a similar way to a slow-roll dark energy during matter domination, and as dark energy becomes more dominant the evolution diverges from the scalar field dynamics. Thawing models act like a cosmological constant for most of cosmic history, only diverging at low-redshift, so they match the early universe predictions of $\Lambda$CDM. This model contains two additional parameters beyond $\Omega_m$, the current day equation of state $w_0$, and $p$, a shape parameter that controls how fast the equation of state thaws. It has Hubble parameter expression
\begin{equation}
        \begin{split}
            \frac{H^2}{H^2_0} & = \Omega_m(1+z)^3 + (1-\Omega_m) \times  \\ & \exp \left( \frac{3(1+w_0)}{\alpha p} \left[1-\left(1-\alpha +\frac{\alpha}{(1+z)^3}\right)^{p/3}\right] \right), 
        \end{split} 
\end{equation}
where $\alpha = \frac{1}{1+b}$, $b = 0.3$.

\subsection*{Dark Energy Transitions}
Dark energy transition models attempt to resolve the Hubble tension by inducing a large step-like response in $H(z)$ at a low redshift via large fluctuations in the equation of state parameter \citep{Mortonson_2009}. This model adds a single parameter $\delta$, where the local observed Hubble constant is $H_0 \approx (1+\delta)\Tilde{H}_0$, and $\Tilde{H}_0$ is the Hubble constant of a reference $\Lambda$CDM model. In this paper the authors suggest a transition at $z_t \approx 0.02$, however our redshift cutoff is above this value, and so we use $z_t = 0.1$ to make use of the low redshift SNe up to $z = 0.1$, as in \cite{Dhawan2020}. This model has Hubble parameter
\begin{equation}
    \frac{H^2}{\Tilde{H}^2_0} = \Tilde{\Omega}_m (1=z)^3 + \left[1+ \frac{2\delta \times \mathcal{S}(z)}{(1-\Tilde{\Omega}_m)\mathcal{S}(0)} \right] (1-\Tilde{\Omega}_m),
\end{equation}
where 
\begin{equation*}
    \mathcal{S}(z)  = \frac{1}{2}\left [ 1 - \tanh\left ( \frac{z-z_t}{\Delta z}\right)\right].
\end{equation*}
The local observed Hubble constant and matter density are then given by 
\begin{align*}
H_0 = \Tilde{H}_0 \sqrt{1+2\delta}, \\
\Omega_m = \Tilde{\Omega}_m \sqrt{1+2\delta}.
\end{align*}

\subsection*{Bimetric Gravity}
Any deviation from a cosmological constant can also be realised by modifications to the equations of GR. This model of Bimetric gravity adds a massive particle to mediate the gravitational force, by having two interacting metrics for space-time instead of just one \citep{deRham2011,mortsell2018}. In this paper we consider the simplest case of Bimetric gravity, including only the linear term of this interaction, which leads to a single additional parameter $B_1$ (where $B_1$ = 0 collapses the theory back down to $\Lambda$CDM). The Hubble parameter expression is given by:
\begin{equation}
    \frac{H^2}{H^2_0} = \frac{\Omega_M(1+z)^3}{2} + \frac{B_0}{6} + \sqrt{\left(\frac{\Omega_M(1+z)^3}{2}+\frac{B_0}{6}\right) ^2+\frac{B_1^2}{3}},
\end{equation}
where 
\begin{equation*}
    B_0 = 3(1-\Omega_M)-B_1^2.
\end{equation*}
\subsection{Likelihood Models}\label{LKmodels}
\begin{figure}
  \begin{minipage}[c]{0.5\textwidth}
  \centering
    \includegraphics{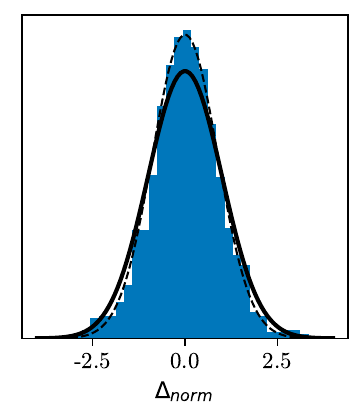}
  \end{minipage}\hfill
  \begin{minipage}[c]{0.5\textwidth}
    \caption{The normalised Hubble residuals of the Pantheon+ data in a best fit F$\Lambda$CDM universe, over-plotted with a standard Gaussian in the solid black line. The residuals are normalised in a way that takes into account covariances, $\Delta_{\text{norm}} = \bm{C}^{-1/2}\Delta$. As the figure shows, the data does not normalise perfectly to the Gaussian distribution, with the data showing a higher peak. The dashed line shows a Gaussian with variance $\sigma^2 = 0.88$.}
  \label{residuals}
  \end{minipage}
\end{figure}

\Cref{residuals} shows the normalised residuals $\bm{C}^{-1/2} \Delta$ for best fit F$\Lambda$CDM parameters. The likelihood function in \cref{gaussianlikelihood} assumes that these residuals will form a unit Gaussian, shown by the continuous black line, but the underlying distribution appears to be slightly tighter. A likelihood function that assigns more accurate weighting to the residuals, so that observations with more or less uncertainty contribute more or less to the likelihood function, will have higher likelihood values, and therefore will be favoured by the evidence.

The discrepancy in \cref{residuals} motivates a series of tests that adjust the form of the likelihood function in \cref{gaussianlikelihood}. Here we apply the Bayesian model comparison methodology not just to the cosmological model, but also to the model of the apparent magnitude scatter. A pair of well motivated models are considered; a generalised Gaussian, and the multivariate Student's t-distribution. \cite{Dainotti} demonstrated a preference for a univariate Student's t-distribution on the normalised residuals, finding a Bayesian Information Criterion of $\Delta_\textrm{BIC} = 134$, which corresponds to an approximate Bayes factor of $67$. We expand upon this by using evidences instead of the BIC, as well as including the non-diagonal elements of the covariance matrix in our analysis.

\subsection*{Generalised Gaussian}
Modelling on 21 \si{cm} cosmology data and artificially added scatter, \cite{Scheutwinkel_2023} found that on data with an unknown ``ground-truth" likelihood function, a generalised Gaussian distribution is a good first order-approximation, and will generally outperform a Gaussian in its Bayesian evidence (unless the underlying scatter is truly Gaussian). In the multivariate case it extends the Gaussian with two additional parameters, scale $A$ and shape $B$ \footnotemark{}, with likelihood function \citep{Pascal_2013}:

\footnotetext{These parameters are often denoted $(\alpha,\beta)$ in the literature, but these characters are reserved for the Tripp light-curve fitting parameters here, so $(A,B)$ are used instead.}
\begin{equation}\label{ABGGD}
    \mathcal{L} = \mathcal{N}(A,B,\bm{C})\exp\left( -\frac{1}{2} \left(\frac{\Delta^T\bm{C}^{-1}\Delta}{A}\right)^{B} \right).
\end{equation}
This has covariance matrix $\bm{M}$ given by:
 \begin{equation} \label{ABcovariance}
     \bm{M} = A \frac{2^{1/B} \Gamma(\frac{N+2}{2B})}{N \Gamma(\frac{N}{2B})} \bm{C}.
 \end{equation}
We extend this distribution even further by introducing a ``skew" parameter $K$, to test the residuals for asymmetry. This combines to give the full distribution:
\begin{equation} \label{ABKGGD}
    \mathcal{L} = \frac{\mathcal{N}(A,B,\bm{C})}{\prod_i (1-K[(A\bm{C})^{-1/2} \Delta]_i)} \exp\left( -\frac{1}{2} (V^T \cdot V)^{B} \right),
\end{equation}
where $V$ are skewed, scaled, normalised residuals:
\begin{equation}
    V_i = \frac{1}{K} \log \left( 1 - K [(A\bm{C})^{-1/2} \Delta]_i \right).
\end{equation}
If any of the residuals are too large ($K (A\bm{C})^{-1/2} \Delta_i > 1$) then the distribution is set to 0, so for higher $K$, large parts of the parameter space are ruled out.

\begin{figure}
 \centering
     \includegraphics{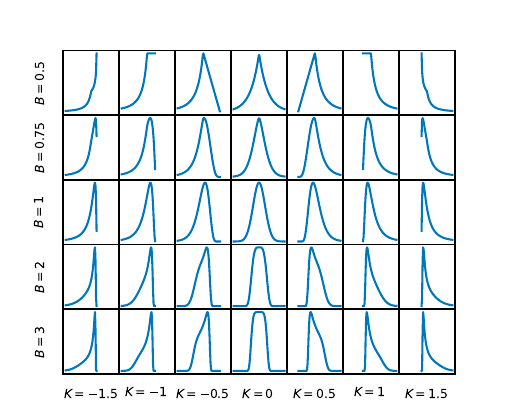}
     \caption{Probability densities of a univariate generalised Gaussian from \cref{ABKGGD} for varying values of $B$ and $K$. The scale parameter and covariance matrix are set to unity, $A = \bm{C} = 1$.}
     \label{generalisedgrid}
 \end{figure}

The full ``Doubly-Generalised" distribution tends back to \cref{ABGGD} as $K \xrightarrow{} 0$, and reverts back to the pure Gaussian in \cref{gaussianlikelihood} for $(A,B,K) = (1,1,0)$. For $A$ and $B$ we take priors $A \sim U[0.5,2]$, $B \sim U(0,3]$, which are sufficiently wide as these bounds extend out to highly non-Gaussian forms. For the skew parameter we take the thin prior $K \sim U[-0.2,0.2]$. Since we have $>1000$ data points we expect to find values at $ > 3\sigma$ even with the best fit parameters, so making $\frac{1}{|K|} < 4$ would have likelihood 0 on almost all of the cosmological parameter space. The univariate case of these distributions are shown for a selection of $B, K$ values in \cref{generalisedgrid}

\subsection*{Student's t}
While the generalised Gaussian is a wide-field first attempt at finding non-gaussianity in the data, we can also try a more specific distribution. The 1D t-distribution is constructed from a Gaussian distribution divided by an independent $\chi^2_\nu$ distribution, so arises as the posterior predictive when working with a Gaussian distribution of unknown variance. Analogously, the multivariate t-distribution shows up as the posterior predictive in a number of marginalisation problems, in particular when estimating the covariance matrix with $\nu$ sums of pairwise deviation products. The process of creating the covariance matrix has uncertainty attached to it, and neglecting this effect by putting in fixed values for the covariance matrix could misrepresent the scatter of the data. This distribution has likelihood function:

\begin{equation} \label{stulikelihood}
 \mathcal{L} =\mathcal{N}(\nu,\bm{C})\left[ 1+ \frac{1}{\nu}(\Delta^T\bm{\Sigma}^{-1}\Delta)\right]^{-(\nu+k)/2},
\end{equation}
where $k$ is the length of the residuals vector, the degrees of freedom $\nu$ is the shape parameter of the distribution, and $\bm{\Sigma}$ is related to the covariance matrix by:
\begin{equation}
    \bm{\Sigma} = \frac{\nu-2}{\nu} \bm{C}.
\end{equation}
For this test we use prior $\nu \sim \log U(2,10^6]$. This bound is chosen as below $\nu = 2$ the t-distribution has no defined variance, and is extended to high $\nu$ since we are running this on a large number of data points. In this case a uniform prior would be inappropriate, since it would over-weigh the high $\nu$ values, $\nu > 10^5$.

\begin{table}
\centering
\caption{Prior distributions for all model parameters in the dark energy models as well as likelihood models tested in this work. For parameters without strong theoretical bounds, wide uniform priors are used.}
\hspace{5ex}  \\

\begin{tabular}{@{\extracolsep{\fill}} l c l}

\hline 
Parameter & Prior & Model \\ [0.5ex] 

\hline 

$M_B$& U[-20,-18] & All \\
$H_0$& U[50,100] & All\\
$\Omega_m$ & U[0,1] & All\\
$w$& U[-2,-2] & $w$CDM (Flat and Curved) \\
$\Omega_k$& U[-0.5,0.5] & $w$CDM and $\Lambda$CDM\\
$w_0$& U[-2,2] & Algebraic thawing\\
$p$& U[-4,4] & Algebraic thawing\\
$B_1$& U[0,6] & Bimetric gravity\\
$\delta w_0$& U[-2,1] & One parameter slow-roll dark energy \\
$\Omega_e$& U[0,0.25] & Growing $\nu$ mass \\
$\Omega_\nu$& U[0,0.4] & Growing $\nu$ mass \\
$\delta$& U[-0.4,0.6] & Dark energy transition  \\ [1ex] 
\hline
$A$& U[0.5,2] & Generalised Gaussian \\
$B$& U(0,3] & Generalised Gaussian \\
$K$& U[-0.2,0.2] & Generalised Gaussian \\
$\nu$& $\log$U(2,$10^6$] & Student's T  \\ [1ex] 
\hline 
\end{tabular}

\label{table:prior}
\end{table}

\renewcommand{\arraystretch}{1.1}
\begin{table*}
\centering
\caption{Parameter estimations and evidences (relative to F$\Lambda$CDM) for each model. All parameter values are quoted as median $\pm$ Bayesian credibility interval. Included also are the odds of each model, given by $\mathcal{Z} \colon \mathcal{Z}_0$.}

\begin{tabular}{@{\extracolsep{\fill}} l c c c c | c c}

\hline 
Model & $H_0$ & $\Omega_m$ & Parameter 2 & Parameter 3 & $\Delta \log \mathcal{Z}$ & Odds \\ [0.5ex] 

\hline 
Flat $\Lambda$ CDM & \err{73.70}{0.98}{1.07} & \err{0.33}{0.02}{0.02} & -- & -- & 0 & -- \\ 
$\Lambda$ CDM & \err{73.65}{1.0}{1.05} & \err{0.32}{0.05}{0.06} & $\Omega_k =$ \err{0.03}{0.13}{0.13} & -- & -1.06 & $<$1:2 \\ 
Flat $w$CDM & \err{73.69}{1.01}{1.07} & \err{0.32}{0.07}{0.06} & $w =$ \err{-0.98}{0.17}{0.15} & -- & -2.33 & $<$1:10 \\
$w$CDM & \err{73.68}{1.07}{1.02} & \err{0.29}{0.07}{0.06} & $\Omega_k =$ \err{0.09}{0.33}{0.21} & $w =$ \err{-1.09}{0.5}{0.29} & -2.33 & $<$1:10 \\
Bimetric Gravity & \err{73.78}{1.0}{1.04} & \err{0.35}{0.03}{0.04} & $B_1 = $ \err{0.57}{0.23}{0.57} & -- & -1.65 & $<$1:5 \\
Slow Roll Inflation & \err{73.69}{1.06}{1.02} & \err{0.33}{0.05}{0.04} & $\delta\ w_0 = $ \err{0.01}{0.16}{0.13} & -- & -2.1 & $<$1:8 \\ 
Algebraic Thawing & \err{73.69}{1.05}{1.03} & \err{0.33}{0.05}{0.05} & $w_0 =$ \err{-1.0}{0.18}{0.14} & $p = $ \err{0.56}{2.71}{2.22} & -2.37 & $<$1:10 \\ 
Growing Neutrino Mass & \err{73.57}{1.01}{1.02} & \err{0.25}{0.07}{0.05} & $\Omega_e =$ \err{0.13}{0.1}{0.07} & $\Omega_\nu=$ \err{0.1}{0.05}{0.1} & -0.75 & $<$1:2 \\
Dark Energy Transition & \err{73.59}{1.48}{1.48} & \err{0.33}{0.04}{0.04} & $\delta =$ \err{0.0}{0.03}{0.03} & -- & -2.57 & $<$1:13 \\  [1ex]  
\hline 

\end{tabular}
\label{parametertable}
\end{table*}
\renewcommand{\arraystretch}{1}
\section{Results}
\label{sec:results}

\subsection{Cosmological Model Comparison}\label{Firstanalysis}
In this section we present the results of the model comparison and parameter estimation on each of the cosmological models.  

\Cref{parametertable} gives the posterior estimates for each cosmological parameter with a Gaussian likelihood, as well as the relative evidences of each model. The parameters are quoted with their median values, and the intervals given are Highest Posterior Density Intervals \citep{HPDI}, a form of Bayesian credibility interval. 

\begin{figure}
    \centering
    \includegraphics[width=.48\textwidth]{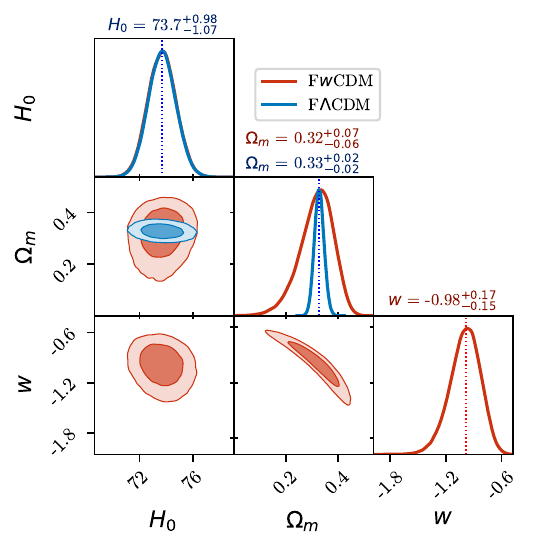}
    \vspace{-0.1cm}
    \caption{The posterior distribution of $H_0$, $\Omega_m$ and $w$ for a Gaussian likelihood. Over plotted are the $H_0$ and $\Omega_m$ posteriors from F$\Lambda$CDM, finding similar values for both values since $w$ is constrained tightly to the null $w = -1$.}
    \label{wCDMTriangle}
\end{figure}

\Cref{wCDMTriangle} shows an example of the 1 and 2D posteriors for F$\Lambda$CDM and F$w$CDM, with the median/interval values over plotted. We can see that both models find a consistent $H_0$ value, showing $w$'s independence from $H_0$, as well as the effect on $\Omega_m$'s constraint by adding the additional parameter.

\begin{figure}
  \includegraphics[width = .48\textwidth, trim = 0 20 0 0]{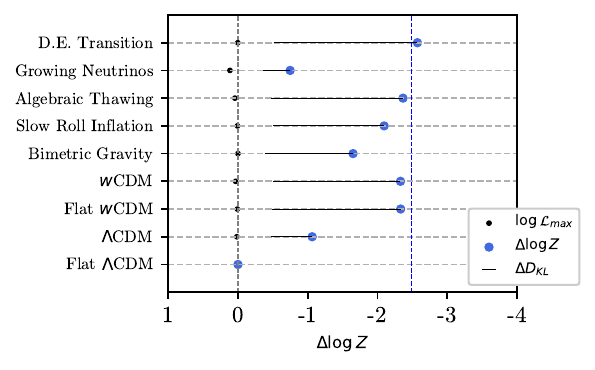}
  \vspace{-0.1cm}
  \caption{The evidences (relative to F$\Lambda$CDM) of each model, run on the set of all SNe with $z>0.023$. The black lines show the $\mathcal{D}_{\textrm{KL}}$ of the models, i.e. the Occam penalty applied to them by the Bayesian evidence. Also shown is the ``Moderate Evidence" cut-off at $\Delta \log \mathcal{Z} = -2.5$, which 1 models fall beyond, a dark energy transition. The black dots show the maximum likelihood point (again relative to F$\Lambda$CDM) achieved by each model.}
  \label{evidence1}
\end{figure}

The relative evidence values are plotted in \cref{evidence1}, where 4 of models are disfavoured at a moderate evidence level (shown by the dashed blue line). The black lines in \cref{evidence1} show the relative $\mathcal{D}_\textrm{KL}$ of each model, the size of the ``Occam Penalty" of each model compared to F$\Lambda$CDM. As well as the evidence values favouring F$\Lambda$CDM, the $\log \mathcal{L}_{max}$ values show that none of the models are able to find a better fit than F$\Lambda$CDM with their additional parameters.

\subsection{Peculiar Velocity Corrections} \label{PVC}
The peculiar velocities (PVs) of the SNe can be up to $300$ \si{km.s^{-1}}, which accounts for $\sim 10\%$ of the total redshift at $z = 0.01$. For higher redshifts, PVs still contribute to the total uncertainty, and the $v_{\text{pec}}$ values used to correct the measured redshifts of the SNe in Pantheon+ are evaluated in \cite{Peterson_2022} and \cite{Carr_2022}. If the process of modelling the large-scale velocity field of the universe out to high redshift has assumed a standard cosmological model, this may have an effect on model comparison, by artificially increasing the fit and evidence for F$\Lambda$CDM. By analysing the Pantheon+ data with and without the peculiar velocity corrections, we can explore whether any possible source of bias is displayed within the relative Bayesian evidences of the models.
This concern is addressed in \cite{Carr_2022}, where parameter estimation is compared on the corrected and uncorrected velocities. They found that the tested parameters ($H_0$ and $w$) do not change appreciably. 
In this study, we present a complementary analysis to the parameter estimation in \citet{Carr_2022}.

\begin{figure}
  \includegraphics[scale = 0.9]{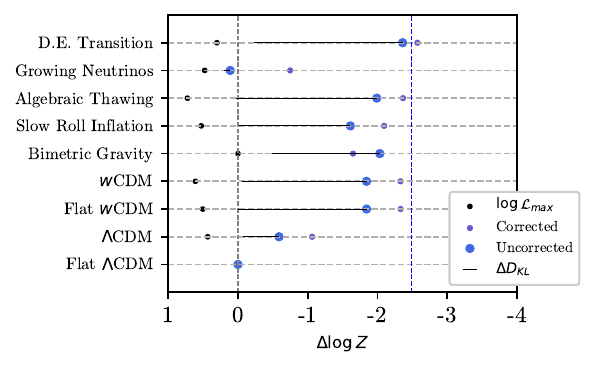}
  \caption{The relative evidences of each model with and without peculiar velocity corrections. Without peculiar velocity corrections no models  fall beyond the moderate evidence line, and all models except Bimetric gravity find a better maximum fit relative to F$\Lambda$CDM than the first analysis \cref{evidence1}, shown by the black $\log \mathcal{L}_\textrm{max}$ dots.}
  \label{evidenceVPEC}
\end{figure}

For this test, instead of using the fiducial case, wherein the redshifts are corrected for the peculiar velocities, we perform model comparison using the CMB frame redshifts with no correction. The relative evidences are shown in \cref{evidenceVPEC}, with the relative evidences from \cref{evidence1} over-plotted. Note that each set of tests (CMB frame and fiducial) are plotted relative to their own F$\Lambda$CDM, as the evidences of the CMB frame tests are much lower than the fiducial. While the distribution of models is roughly the same for both, in the uncorrected data all models perform slightly better as compared to F$\Lambda$CDM, with $|\Delta \log \mathcal{Z}| \sim 0.5$ smaller on average.

\subsection{Alternate Likelihood Models} \label{altlike}
The initial test is repeated, taking the same Hubble Flow and Calibrator SNe set, but now through each of the alternate likelihood models described in \cref{LKmodels}. For the generalised Gaussian, we run a test with each combination of the parameters free/fixed, giving seven alternate forms of the likelihood function, and these tests are referred to by their free parameters: $A$,$B$,$K$,$AB$,$AK$,$BK$,$ABK$. This gives nine total likelihoods, along with the regular Gaussian and the Student's t. The evidences of each cosmology/likelihood model pair are given in \cref{margtable}, and all cosmology-marginalised parameter values are given in \cref{scatterparametertable}. 

\begin{figure}
     \centering
     \includegraphics{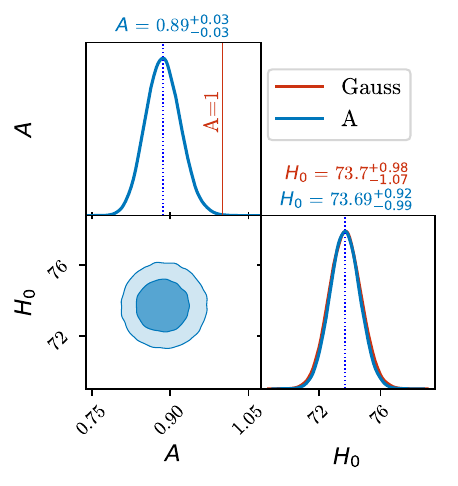}
     \caption{The posterior for $A$ and $H_0$ in A-scaled F$\Lambda$CDM (blue), compared with the fiducial analysis (orange). $A$ takes on a distinctly non-null value, and as a result the $H_0$ estimate is slightly tighter.}
     \label{Agausscomp}
 \end{figure}

For the set of generalised Gaussian models, the results of the model comparison in \cref{margevscatter} can be explained by the estimates of each of the parameters $A$, $B$, and $K$, as well as their relationships to each other. As shown in \cref{Agausscomp} $A$ is centered on $0.89 \pm 0.03$, finding a distinctly non-null value, and this results in a better fit. On the other hand, both $B$ and $K$ are tightly constrained to their null values: $ B = 1.02 \pm 0.005$ and $K = 0.02 \pm 0.02$. This means that the $B$, $K$, and $BK$ tests all underperform the Gaussian; they introduce an unnecessary parameter and are punished according to Occam's razor.

For a multivariate distribution with $N = 1448$, the covariance of the generalised Gaussian, given in \cref{ABcovariance}, is extremely sensitive to $B$, leading in this case to a very tight posterior centred on $B=1$. This can also be seen when considering that for a normal distribution, the distance term $(\Delta^T \bm{C}^{-1}\Delta)^B \sim N^B$, so any non-unity $B$ in the exponent has an exponential effect on the size of this term. As well as this, $A$ and $B$ have a degenerate effect on the scale of the covariance, which means that in the $AB$ test this degeneracy breaks the strong $A = 0.89$ preference that had benefited the model. This also results in the full $ABK$ model finding a disfavoured evidence, in combination with the data finding a null value of the skew.

Finally, $A$ and $K$ are almost independent in their effect on the model comparison. This is shown in the ratio of their evidences:
\begin{equation}
    \frac{\mathcal{Z}_K}{\mathcal{Z}_\text{Gaussian}} \approx \frac{\mathcal{Z}_{AK}}{\mathcal{Z}_A}.
\end{equation}
This result is reassuring: $A$ is analogous to the covariances being uniformly smaller/larger, which is independent from the role of K in skewing the distribution (especially since it skews the residuals after normalisation), so K's effect on the evidence shouldn't change based on whether $A$ is free or fixed.

\begin{figure}
    \centering
    \includegraphics{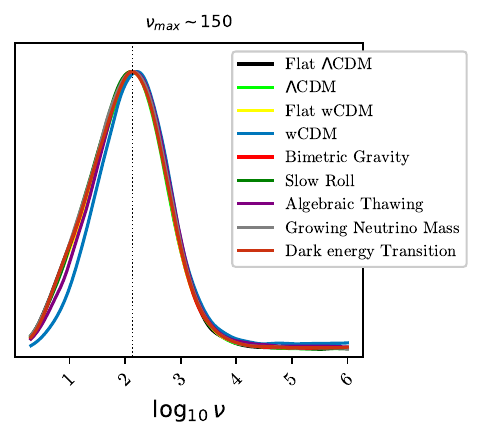}
    \caption{The 1D posterior distribution of $\nu$ on each cosmological model. The data shows a clear preference for lower $\nu$ values, with the posterior dying off where the distribution limits towards a Gaussian.}
    \label{numarg}
\end{figure}

The t-distribution also find an improvement over the Gaussian of $\Delta \log \mathcal{Z} = +2.46$, or a Bayes factor of $B = e^{2.46} = 11.7$. \Cref{numarg} shows the 1D posterior for $\nu$ on each cosmological model. If the posterior was one sided and increasing towards higher $\nu$, this would suggest that the more Gaussian the distribution, the better the fit, but while the value of $\nu$ is not tightly constrained, it peaks at $\nu \sim 150$ on each model. The posterior and evidence show a preference for a non-Gaussian form.
 
\subsection{Model Marginalisation} \label{modelmarg}

In the case of having both a selection of cosmological models and a selection of scatter models, we can marginalise out one or both of these choices to obtain either: \textbf{a)}  Cosmology independent evidences for each scatter model or \textbf{b)} Scatter-model independent evidences for each cosmology.
Labeling the cosmological model as $\mathcal{M}_c$ and scatter model $\mathcal{M}_s$, we marginalise the evidence (as in \cref{Marginalisation}):
\begin{equation} \label{modelmarg2}
    p(d|\mathcal{M}_s) = \sum_{\mathcal{M}_c} p(d|\mathcal{M}_c,\mathcal{M}_s)p(\mathcal{M}_c|\mathcal{M}_s),
\end{equation}
where since $p(\mathcal{M}_c|\mathcal{M}_s)$ is not conditioned on the data it takes the prior values, i.e. is uniform.

\begin{figure}
  \includegraphics{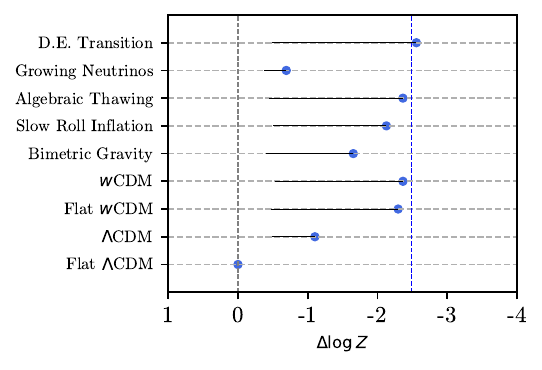}

  \caption{The relative evidences of each cosmology, marginalised over the scatter model. F$\Lambda$CDM has the highest evidence, providing similar results to the Gaussian analysis (\cref{evidence1}). The black lines show the joint $\mathcal{D}_\textrm{KL}$ of the scatter model/parameters.}
  \label{margevcosmo}
\end{figure}

\begin{figure}
  \includegraphics{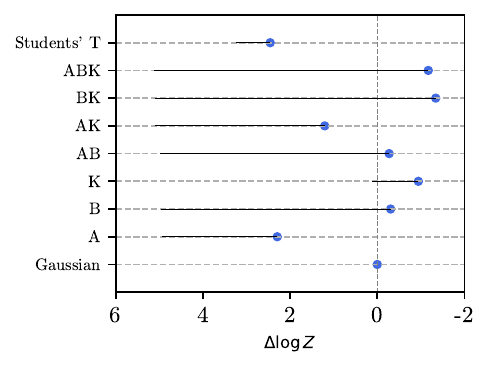}
  
  \caption{The relative evidences of each scatter model, marginalised over the cosmological model. Two likelihood functions, the $A$-scaled Gaussian and the Student's t-distribution are favoured over the Gaussian. The black lines show the joint $\mathcal{D}_\textrm{KL}$ of the cosmological model/parameters.}
  \label{margevscatter}
\end{figure}

This equation also holds with ``$c$" and ``$s$" swapped to marginalise out the scatter model on the cosmological evidences. The results of these marginalisations are shown in \cref{margevcosmo,margevscatter}. Since all of the summing in \cref{modelmarg2} takes place outside of the $\log$, the results are weighted very strongly towards high evidence models, i.e the marginalised likelihood comparison doesn't vary appreciably from the likelihood comparison for F$\Lambda$CDM.

We also calculate the total $\mathcal{D}_\textrm{KL}$ for each cosmology/scatter model, by considering the $\mathcal{D}_\textrm{KL}$ of the joint distribution over model and parameter space:
\begin{equation}
    p(\theta , \mathcal{M}_s|d,\mathcal{M}_c) = p(\theta|d,\mathcal{M}_c,\mathcal{M}_s) p(\mathcal{M}_s|d,\mathcal{M}_c).
\end{equation}
From \cite{kroupa2023}, the joint $\mathcal{D}_\textrm{KL}$ can be broken down as:
\begin{equation}
	\begin{split}
        \mathcal{D}_\textrm{KL}(\theta,\mathcal{M}_s|\mathcal{M}_c) & = \mathcal{D}_\textrm{KL}(\mathcal{M}_s|\mathcal{M}_c) + \\ & \sum_{\mathcal{M}_s} p(\mathcal{M}_s|d,\mathcal{M}_c) \mathcal{D}_\textrm{KL}(\theta| \mathcal{M}_s, \mathcal{M}_c),
	\end{split}
\end{equation}
i.e. the posterior averaged $\mathcal{D}_\textrm{KL}$ from each scatter model, as well as the $\mathcal{D}_\textrm{KL}$ of the (discrete) distribution of scatter models.
These $\mathcal{D}_\textrm{KL}$ are plotted on the marginalised evidence plots \cref{margevcosmo,margevscatter}. For the cosmological models, each model finds the same fit as F$\Lambda$CDM, and the $\mathcal{D}_\textrm{KL}$s line up at $\Delta \log \mathcal{Z} \approx -0.5$ just as in the Gaussian case.

Fully marginalising over both cosmology and scatter models should give the best estimate of $H_0$ that SNe are able to offer, using the data to the fullest extent. Marginalising over models gives:
\begin{equation} \label{Marginalisation}
    p(H_0|d) = \sum_{\mathcal{M}_c,\mathcal{M}_s} p(H_0|d,\mathcal{M}_c,\mathcal{M}_s) p(\mathcal{M}_c,\mathcal{M}_s|d).
\end{equation}
Over the marginalised posterior, we find a value of $H_0 = 73.67\pm0.99$ \si{km.s^{-1}.Mpc^{-1}}, in agreement with the results of the SH$_0$ES collaboration \cite{Riess2022}, $H_0 = 73.04 \pm 1.04$ 
 \si{km.s^{-1}.Mpc^{-1}}, although with slightly tighter constraints owing to the more general treatment of the distribution of the residuals. 

All other parameters are similarly marginalised, and given in \cref{margparametertable,scatterparametertable}, along with the marginalised evidences and odds.
\begin{table*}
\centering
\caption{Evidences for each cosmological/likelihood pair model. The marginalised evidences are given for a cosmology/scatter independent model comparison, as described in \cref{modelmarg}.}

\begin{tabular}{@{\extracolsep{\fill}} l c c c c c c c c c| c }

\hline 
Model & Gaussian & $A$ & B & K & $AB$ & $AK$ & $BK$ & $ABK$ & Student's T & \textbf{Scatter Independent} \\ [0.5ex] 
\hline 
Flat $\Lambda$CDM & 811.2 & 813.5 & 810.9 & 810.3 & 811.0 & 812.5 & 809.9 & 810.1 & 813.7 & 812.4 \\
$\Lambda$CDM & 810.2 & 812.5 & 809.8 & 809.2 & 809.8 & 811.3 & 808.8 & 808.9 & 812.6 & 811.3 \\
Flat $w$CDM & 808.9 & 811.2 & 808.7 & 808.1 & 808.6 & 810.2 & 807.6 & 807.8 & 811.4 & 810.1 \\
$w$CDM & 808.9 & 811.2 & 808.5 & 808.1 & 808.7 & 810.1 & 807.6 & 807.7 & 811.3 & 810.0 \\
Bimetric Gravity & 809.6 & 811.8 & 809.3 & 808.7 & 809.3 & 810.8 & 808.2 & 808.4 & 812.1 & 810.7 \\
Slow Roll Inflation & 809.1 & 811.4 & 808.8 & 808.2 & 808.8 & 810.3 & 807.8 & 808.0 & 811.5 & 810.3 \\
Algebraic Thawing & 808.9 & 811.2 & 808.5 & 808.1 & 808.6 & 810.2 & 807.6 & 807.8 & 811.3 & 810.0 \\
Growing Neutrino Mass & 810.5 & 812.9 & 810.2 & 809.6 & 810.3 & 811.7 & 809.1 & 809.3 & 813.0 & 811.7 \\
Dark Energy Transition & 808.7 & 811.0 & 808.8 & 807.7 & 808.3 & 809.8 & 807.4 & 807.4 & 811.1 & 809.8 \\ \hline
\textbf{Cosmology Independent} & 810.0 & 812.3 & 809.7 & 809.0 & 809.7 & 811.2 & 808.6 & 808.8 & 812.4 &  -- \\
\hline
\end{tabular}
\label{margtable}
\end{table*}

\subsection{Transitional and Changing Absolute Magnitudes} \label{MBTRANS}
\begin{figure}
    \centering
        \includegraphics{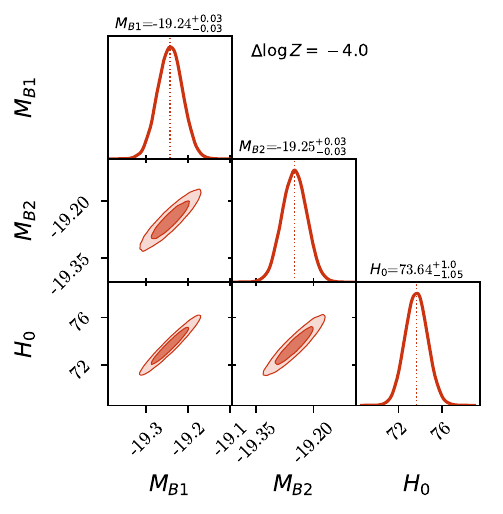}
        \caption{The posterior for an $M_B$ transition model, showing the near and far magnitudes (split about redshift z =0.1). Both values are tightly centred on the Cepheid calibrated value.}
        \label{doublemb}
\end{figure}

\begin{figure}
    \centering
    \includegraphics{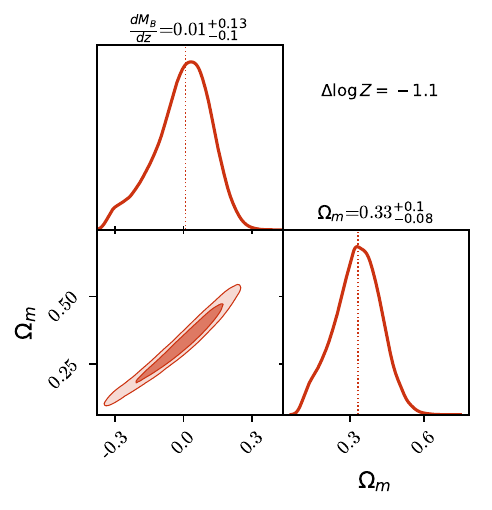}
         \caption{The posterior for a linear $M_B$ model. The figure shows a degeneracy with $\Omega_m$, and approximately triples the size of its error bounds compared to F$\Lambda$CDM.}
         \label{linearMB}
\end{figure}

The dark energy transition model attempts to resolve the Hubble tension by inducing a step-like transition in $H_0$ at low redshift via a spike in $w(z)$. This model is highly phenomenological, and the required change in $w(z)$ is theoretically permitted, but not theoretically motivated. Since $M_B$ and $H_0$ have a degenerate effect on our Hubble residuals, a step-like change in $H_0$ can also be realised by a changing absolute magnitude. We test two additional models, both working in a F$\Lambda$CDM cosmology:

\begin{equation}
\label{doubleMBequation}
M_B =
\begin{cases}
M_{B1} & \text{if } z \leq 0.1 \\

M_{B2} & \text{if } z > 0.1
\end{cases}
\end{equation}
\begin{equation}
\label{linearMBequation}
    M_B = M_{B1} + \frac{z}{z_{\textrm{max}}}\times(M_{B2} - M_{B1}).
\end{equation}
\Cref{doubleMBequation} mimics the effect of the dark energy transition model on $H_0$, a sharp step in $M_B$ at $z = 0.1$. This adds an additional step to our distance ladder; the Cepheid distances constrain $M_{B1}$ independent of $H_0$, so that the ``near" supernovae can constrain $H_0$, which is used to constrain $M_{B2}$. 
\Cref{linearMBequation} is a first order test for an evolving $M_B$, by including a steady $M_B - z$ gradient. \cite{Kistler_2013}, \cite{Meng2011} both suggest that SNe~Ia evolve with metallicity (which can change the amount of radioactive material in the explosion, and hence the luminosity of the SN~Ia). Since SNe~Ia measure dark energy via relative distance measurements, any intrinsic evolution with cosmic time must be accounted for. Both models are run in an F$\Lambda$CDM cosmology. The triangle plots for both models are given in \cref{doublemb,linearMB}, as well the evidence values relative to F$\Lambda$CDM run on the same dataset.

Both dynamic $M_B$ models come out with lower evidence than F$\Lambda$CDM, and similarly to the dark energy models have their additional parameters constrained to their null values. The data shows no support for a dynamic $M_B$. 

The transition model is ruled out at a higher level than the dark energy transition model, $\Delta \log \mathcal{Z} = -4.0$ relative to F$\Lambda$CDM. This model has lower evidence than the dark energy transition model, likely because the prior range of $M_{B1} \in [-20,-18]$ has a much larger effect than the dark energy transition parameter $\delta \in [-0.4,0.6]$.

The lack of an $M_B$ transition shown in \cref{doublemb} can be attributed to the same reason a dark energy transition is not favoured by the evidence; the effect may not occur at $z_t = 0.1$. However, due to the high peculiar velocity error in the lower end of this data set, we are unable to test the suggested lower value of $z_t = 0.02$.
For the linear model, the $M_B$ gradient isn't tightly constrained, as its effect is reduced on lower redshifts (where much of the data lies), and shows some degeneracy with $\Omega_m$ in \cref{linearMB}. This suggests that neglecting the effects of a magnitude gradient could lead to over or underestimating the matter density. However, SNe~Ia evolution is likely to be reflected in the light curve of the SNe, which is precisely what the Pantheon+ data have been calibrated to standardise. Therefore, the lack of an effect here suggests that the $M_B$ corrections used in the Pantheon+ catalogue have already corrected for a first order $M_B$ gradient. The negative $\Delta \log \mathcal{Z} = -1.1$ of this model reflects the lack of improvement of fit, and the weakness of the evidence difference reflects the weakness of the constraint on the $M_B$ gradient (as in the model comparison discussion).

Importantly for both of these models the $H_0$ estimation remains the same, and does not vary drastically towards the Planck value (although the uncertainty in $H_0$ is doubled by the inclusion of the $M_{B2}$ parameter). This paper finds no evidence to suggest that the Hubble Tension can be resolved via a step-like or linear $M_B - z$ relationship.

\section{Discussion}\label{sec:discussion}

\renewcommand{\arraystretch}{1.1}
\begin{table*}
\centering
\caption{Parameter estimations and evidences for each cosmological model, now marginalised over the scatter model. All parameter values are quoted as median $\pm$ Bayesian credibility interval, and again the (marginalised) odds and evidences of each model are listed, with odds given as $\mathcal{Z}\colon \mathcal{Z}_0$.}

\begin{tabular}{@{\extracolsep{\fill}} l c c c c | c c}

\hline 
Model & $H_0$ & $\Omega_m$ & Parameter 2 & Parameter 3 & $\Delta \log \mathcal{Z}$ & Odds \\ [0.5ex] 
\hline 
Flat $\Lambda$ CDM & \err{73.7}{0.93}{1.0}& \err{0.33}{0.02}{0.02}& -- & -- & 0 & -- \\ 
$\Lambda$ CDM & \err{73.67}{0.96}{0.99}& \err{0.32}{0.05}{0.05}& $\Omega_k =$ \err{0.03}{0.12}{0.12}& -- & -1.11 & $<$1:3 \\ 
Flat $w$CDM & \err{73.69}{0.97}{0.99}& \err{0.32}{0.06}{0.06}& $w =$\err{-0.98}{0.16}{0.14}& -- & -2.30 & $<$1:10 \\
$w$CDM &\err{73.71}{0.98}{1.0}& \err{0.29}{0.07}{0.06}& $\Omega_k =$ \err{0.09}{0.32}{0.22}& $w =$\err{-1.09}{0.52}{0.28} & -2.37 & $<$1:10 \\
Bimetric Gravity & \err{73.77}{0.97}{0.97}& \err{0.35}{0.03}{0.04}& $B_1 =$ \err{0.55}{0.22}{0.55}& -- & -1.65 & $<$1:5 \\
Slow Roll Inflation & \err{73.69}{0.96}{1.01}& \err{0.33}{0.05}{0.04}& $\delta w_0 =$\err{0.01}{0.15}{0.13}& -- & -2.13 & $<$1:8 \\ 
Algebraic Thawing & \err{73.7}{0.97}{1.0}& \err{0.33}{0.05}{0.05}& $w_0 =$ \err{-1.0}{0.16}{0.14}& $p = $ \err{0.55}{2.54}{2.4} & -2.36 & $<$1:10 \\ 
Growing Neutrino Mass &  \err{73.58}{0.94}{1.0}& \err{0.26}{0.07}{0.05}& $\Omega_e = $ \err{0.13}{0.11}{0.06}& $\Omega_\nu =$ \err{0.1}{0.04}{0.1} & -0.69 & $<$1:2 \\
Dark Energy Transition & \err{73.61}{1.43}{1.37}& \err{0.33}{0.03}{0.04}& $\delta = $ \err{0.00}{0.03}{0.03}& -- & -2.56 & $<$1:13 \\  [1ex]  
\hline 

\end{tabular}
\label{margparametertable}
\end{table*}
\renewcommand{\arraystretch}{1}

\subsection{Alternate Likelihood Distributions}
From the forms of the various likelihood functions in \cref{gaussianlikelihood,ABGGD,,ABKGGD,,stulikelihood}, each distribution is (naturally) peaked by the minimising the residuals $\Delta^T \bm{C}^{-1} \Delta$ (except for the skewed distributions $K\neq 0$ which have their peak shifted only by a small factor). As a result the choice of likelihood function doesn't change the central value of the parameter estimation, and therefore the conclusions of the model comparison don't change over different scatter models. Additionally, since all alternate cosmologies are converging to their  F$\Lambda$CDM limit, scatter model parameters are not sensitive to the specific cosmological model. This leads to an independence in this analysis between the scatter and cosmological models.

The two scatter models with the greatest evidence - a scaled Gaussian with only $A$ as a free parameter and the student's-$t$ distribution - appear to succeed for unrelated reasons, suggesting there could be more than one factor contributing to the improvement of fit in each case.

It is clear why the A-scaled Gaussian succeeds; from \cref{residuals} the Hubble residuals are packed more tightly than the Pantheon+ covariance predicts, and therefore tightening the distribution with $A = 0.89$ allows the smaller residuals to contribute more to the likelihood and find a better fit. \Cref{Agausscomp} also shows that uniformly scaling the Pantheon+ covariances down makes the uncertainty on other model parameters smaller, a natural result considering that for a uniform prior, the likelihood and posterior are proportional. A-scaled F$\Lambda$CDM achieves the tightest $H_0$ of any model, at $\pm 0.93$. This, more clearly than any marginalised result, displays the usefulness of treating the residuals properly in achieving better parameter estimates. As motivation to look into the root of this result; if we had an a priori reason to insert this scaled covariance matrix (and ignore all of the other tested likelihoods) then we would gain the full benefit of the lower variances and we could, in good faith, report this even tighter $H_0$ estimate.

The t-distribution is less immediately interpreted. In a full Bayesian treatment, instead of using the Pantheon+ covariance matrix as it is, we would incorporate the observations that led to the creation of this covariance matrix (such as SNe simulations), as well as a prior for the covariance matrix, into our analysis. This would allow us to marginalise out the true covariance and construct a posterior predictive, i.e. a likelihood that is conditioned on the previously observed SNe data. In the case that the prior for the covariance matrix was an uninformative inverse-Wishart distribution, the correct likelihood for the data would be a multivariate t-distribution, where the shape parameter $\Sigma$ is the sample sum of squares of the input data. In this case the degrees of freedom $\nu = n-k$, where $n$ is the number of observations used to construct the covariance matrix, and $k$ is the size of the covariance matrix (1448). 

While a 1D t-distribution with degrees of freedom $\nu \geq 10$ is indistinguishable from a Gaussian by eye, in the multivariate case much higher $\nu$ values are relevant, due to the higher dimensionality. This can be seen in the $\nu+k$ term in the exponent of \cref{stulikelihood}. Over 1448 data points, this difference can be resolved from the standard Gaussian distribution, and a statistical preference for $\nu \sim 150$ can be identified. The construction described above is not exactly how this inference runs, but treating the covariance matrix as a sample covariance matrix estimated as a sum of deviations, and accounting for the uncertainty in this via a t-distribution, is preferred by the evidence, in a cosmology independent way, by a Bayes factor of $ e^{2.46} \approx 12$. Although the bound is not very tight, the posterior on $\nu$ says that the covariance matrix was effectively estimated with $ n = k+\nu \sim 2000$ observations. For a full derivation of these results, see Chapter 3.6 of \cite{BDA}.

\renewcommand{\arraystretch}{1.1}
\begin{table*}
\centering
\caption{Parameter estimations and evidences for each scatter model, marginalised over the cosmological model. All parameter values are quoted as median $\pm$ Bayesian credibility interval, and the marginalised odds and evidences of each scatter model are listed, with odds given as $\mathcal{Z}\colon \mathcal{Z}_0$.}
\begin{tabular}{@{\extracolsep{\fill}} l c c c | c c}

\hline 
Model & Parameter 1 & Parameter 2 & Parameter 3 & $\Delta \log \mathcal{Z}$ & Odds \\ [0.5ex] 
\hline 
Gaussian & -- & -- & -- & 0 & -- \\
$A$ & $A$ = \err{0.89}{0.03}{0.03}& -- & -- & +2.29 & >10:1\\
$B$ & $B$=\err{1.02}{0.005}{0.005}& -- & -- & -0.31 & <1:1.3\\
$K$ & $K$=\err{0.02}{0.02}{0.02}& -- & -- &-0.94 & <1:3\\
$AB$ & $A$=\err{1.31}{0.68}{0.27}& $B$=\err{1.05}{0.06}{0.03}& -- &-0.27 & <1:1.3\\
$AK$& $A$=\err{0.89}{0.03}{0.03}& $K$=\err{0.02}{0.02}{0.02}& -- & +1.21 & >3:1\\
$BK$& $B$=\err{1.02}{0.005}{0.005}& $K$=\err{0.02}{0.02}{0.02}& -- & -1.34 & <1:4\\
$ABK$& $A$=\err{1.42}{0.58}{0.26}& $B$=\err{1.06}{0.05}{0.03}& $K$=\err{0.02}{0.02}{0.02}& -1.17 & <1:3 \\
Student's t & $\log_{10}(\nu)$=\err{2.09}{0.72}{0.77}& -- & -- & +2.46 & >12:1\\
  [1ex]  
\hline

\end{tabular}
\label{scatterparametertable}
\end{table*}
\renewcommand{\arraystretch}{1}

\subsection{Model Comparison}\label{modelcompdiss}
From \cref{margevcosmo}, the Occam penalty approximation \cref{occampenaltyapprox} holds; the difference between where the black $\mathcal{D}_\textrm{KL}$ lines end ($\langle \log \mathcal{L}\rangle_\mathcal{P}$) and the black $\log \mathcal{L}_{\textrm{max}}$ dots is in all cases $\approx 1/2$, and each model has $\Delta \hat{d} \approx 1$.

In both the Gaussian and the marginalised analysis (\cref{parametertable,margparametertable}) all parameters lie within $1\sigma$ of their null value, i.e. the value that turns the theory back into F$\Lambda$CDM, and therefore all models find the same fit as F$\Lambda$CDM. The alternate dark energy models do not achieve a significantly better fit on the data, and due to the Occam penalty on their expanded prior space have lower evidence than F$\Lambda$CDM in all cases, where the scale of the evidence difference is determined solely by the $\mathcal{D}_\textrm{KL}$.
 
With some simplifying assumptions we can calculate the theoretical $\Delta \log \mathcal{Z}$ of a future model comparison. Consider a one parameter expansion to F$\Lambda$CDM, where the posterior of the additional parameter is still centred around its null value (so that $\Delta \log \mathcal{L}_{\textrm{max}} \approx 0$), and the additional parameter $\theta^*$ is approximately independent of $\Omega_M$ (so that $\Delta \mathcal{D}_\textrm{KL} \approx \mathcal{D}_\textrm{KL}(\theta^*)$). For the sake of simplicity, give this parameter a uniform prior width $b$ and Gaussian posterior with variance $\sigma^2$. Additionally assume that the prior is sufficiently wide, so that $\mathcal{D}_\textrm{KL}$ can be effectively evaluated over $\pm\infty$. Calculating $\mathcal{D}_\textrm{KL}(\theta^*)$ analytically, we find:

\begin{equation}
\int P(\theta^*)\log(P(\theta^*)/\pi (\theta^*)) d \theta^* = \log(b/\sigma) -\frac{1}{2}\log (2\pi) -\frac{1}{2}.
\end{equation}
With $\Delta \hat{d} = 1$ this provides a difference in evidence of
\begin{equation} \label{approxdeltalogz}
	\begin{split}
		\Delta \log \mathcal{Z} & = (\langle \log \mathcal{L}\rangle_\mathcal{P}-\mathcal{D}_\textrm{KL})_1-(\langle \log \mathcal{L}\rangle_\mathcal{P}-\mathcal{D}_\textrm{KL})_0 \\ & \approx - \mathcal{D}_\textrm{KL}(\theta*) - \frac{1}{2} = \frac{1}{2}\log (2\pi)-\log(b/\sigma).
	\end{split}
\end{equation}
So for the well behaved, one-parameter extensions that centre around F$\Lambda$CDM (F$w$CDM, $\Lambda$CDM, Slow Roll) the Bayes factor has a clear relationship to the constraint on the additional parameter; $B_{01} \propto \sigma^{-1}$. While this result is only exact in a highly idealized case, as a heuristic it informs us how much more precise future surveys need to become to rule out parameters like curvature with SNe alone. The strength of our belief in discarding an extension to our base model is directly proportional to how tightly we can constrain its added parameters to 0.

\subsection{Peculiar Velocity Corrections}
From the evidence diagram \cref{evidenceVPEC} the inference with the CMB frame redshifts does not distinguish between the dark energy models as well as inference after peculiar velocity correction are added to the redshifts.
For each model, the Bayesian evidence for the data with peculiar velocity corrections is significantly greater, e.g. by $\sim$ 10 compared to the data without corrections. The smaller evidence differences in the uncorrected analysis could be caused by the data having less constraining power, due to the additional noise. From \cref{approxdeltalogz} if additional parameters cannot be constrained as tightly, then evidence differences will be smaller.

\begin{figure}
  \includegraphics{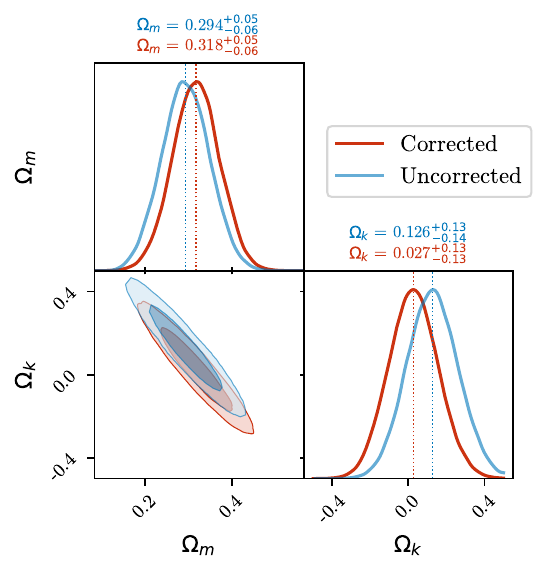}
  \caption{The scatter added by removing peculiar velocity corrections appears to be better fit by a non-zero curvature, as shown in the blue posterior. This is the reason that the evidence for curved $\Lambda$CDM shifts closer to the default cosmology when run on the uncorrected data.}
  \label{vpeccomp}
\end{figure}

In the first analysis (\cref{evidence1}), the black $\log \mathcal{L}_{\textrm{max}}$ dots line up with F$\Lambda$CDM; the alternate models provide no improvement in fit. As discussed before, because all alternate models contain F$\Lambda$CDM this implies that their additional parameters are all being constrained around their null values. In the case of the uncorrected data however, all models except for Bimetric gravity are now achieving a slightly better fit, and the evidence difference is no longer just the penalty terms. The additional scatter that is added by removing peculiar velocity corrections is fit well by all models except Bimetric gravity, in a way that the scatter in the normal analysis does not favour any model over F$\Lambda$CDM. As a result, the growing neutrino mass model is now equally favoured by the evidence, at $\Delta\log\mathcal{Z} = +0.11$. In order for a model to achieve an improvement in fit it needs to have one of its parameter's stray from the null value, displayed most clearly with curvature in $\Lambda$CDM. \Cref{vpeccomp} shows the posterior of $\Lambda$CDM with and without velocity corrections, where there is a distinct preference for a non-zero curvature in the uncorrected analysis. 
 
\section{Conclusions}\label{sec:conclusion}
In this paper we have considered nine dark energy models, covering $\Lambda$CDM and its extensions, dynamical dark energy theories, and a theory of modified gravity. We find that Flat $\Lambda$CDM is preferred by the data over all models (\cref{margevcosmo}), and this preference holds independent of the form of the likelihood function.

When removing peculiar velocity redshift corrections, the growing neutrino mass model is favoured ahead of F$\Lambda$CDM (although not at any significance), and on all other models the degree of the model comparison is weakened without the corrections. This could simply be caused by a lack of constraining power on the uncorrected data, however the values of the $\Delta \log \mathcal{L}_{\textrm{max}}$ suggest that the model have a different fit on the uncorrected redshifts. This could suggest that the assumed F$\Lambda$CDM model used in velocity field reconstruction introduces a slight bias towards F$\Lambda$CDM when performing model comparison.

On the handful of alternate likelihood functions tested, we find evidence to suggest that the covariance matrix should be uniformly scaled down, or equal evidence that the residuals should be modelled with a Student's t-distribution. The success of these two distributions appear to contradict one-another; the t-distribution succeeds because it correctly models the additional uncertainty in the estimation of the covariance matrix, whereas the A-only test succeeds by uniformly reducing the uncertainty. We believe that a more precise formulation of the t-distribution on this data could clarify the emergence of this non-Gaussian result, and reconcile this apparent contradiction.

As a result of this treatment of the residuals on a more general class of likelihood functions, the uncertainty on the Hubble constant is reduced by $5\%$. We have tightened the $H_0$ constraint with no new data by more appropriately modelling the scatter, and regardless of the increase in precision, the scatter-independent parameter estimates of \cref{margparametertable} are a more honest reflection of the data, as they marginalise out our uncertainty about the form of the likelihood function. Future work into likelihood-free inference could result in being able to constraint $H_0$, and other parameters, even further. This would improve not only current estimation and model comparison, but set us up well for the next generation of future telescopes, which are expected to increase the size of SNe catalogues by an order of magnitude \citep{rose2021}. 

It's worth stressing that all of the conclusions of this paper are based on SNe~Ia alone, achieving precision cosmology results without high-redshift constraints, and evaluating the cosmologies through only a single mechanism (luminosity distances). Incorporating other refined probes (e.g. BAOs, CMB) with this improved analysis of SNe could lead to a similar increase in precision to that found in this paper, on a much broader set of parameters and cosmological tests.

\section*{Acknowledgements}

This work was performed using the Cambridge Service for Data Driven Discovery (CSD3), part of which is operated by the University of Cambridge Research Computing on behalf of the STFC DiRAC HPC Facility (www.dirac.ac.uk). The DiRAC component of CSD3 was funded by BEIS capital funding via STFC capital grants ST/P002307/1 and ST/R002452/1 and STFC operations grant ST/R00689X/1. DiRAC is part of the National e-Infrastructure.


\section*{Data Availability}
The Pantheon+ dataset is publicly available on the Pantheon+ repository, at 
https://github.com/PantheonPlusSH0ES/DataRelease/

The nested sampling chains underlying this article are available on Zenodo, at https://doi.org/10.5281/zenodo.10026539



\bibliographystyle{mnras}
\bibliography{nongaussianityinSNE} 



\bsp	

\label{lastpage}

\end{document}